\documentclass[referee,useAMS,usenatbib]{mn2e}
\usepackage{amsmath,natbib}
\usepackage{graphicx,amssymb}
\def\bea{\begin{eqnarray}}
\def\eea{\end{eqnarray}}
\def\vec#1{\ifmmode\mathchoice{\mbox{\boldmath$\displaystyle#1$}}
{\mbox{\boldmath$\textstyle#1$}} {\mbox{\boldmath$\scriptstyle#1$}}
{\mbox{\boldmath$\scriptscriptstyle#1$}}\else
\hbox{\boldmath$\textstyle#1$}\fi}

\begin{document}

\title[Microlensing with the
space interferometer Radioastron] {Microlensing with the space
interferometer Radioastron}
\author[A.F.~Zakharov] {A.F.~Zakharov$^{1,2,3,4}$,\thanks{E-mail:
zakharov@itep.ru}
\\
$^1$ National Astronomical Observatories of Chinese Academy of
Sciences, Beijing 100012, China\\
 $^2$Russian Scientific Centre --
Institute of Theoretical and Experimental Physics, 117259, Moscow, Russia\\
$^3$Astro Space Centre of Lebedev Physics Institute, Moscow, Russia\\
$^4$BLTP, Joint Institute for Nuclear Research, Dubna, Russia}

\maketitle

\begin{abstract}

It is well-known that gravitational lensing is a powerful tool to
investigate matter distributions including DM. Typical angular
distances between images and typical time scales depend on
gravitational lens masses. For microlensing case angular distances
between images or typical astrometric shifts due to microlensing are
about $10^{-5}-10^{-6}~\mu as$. Such an angular resolution will be
reached with the space space--ground interferometer Radioastron. The
basic targets for microlensing searches should be bright point-like
radio sources at cosmological distances. In this case, an analysis
of their variability and a solid determination of microlensing could
lead to an estimation of their cosmological mass density, moreover,
in this case one could not exclude a possibility that non-baryonic
dark matter also form microlenses if the corresponding optical depth
will be high enough. To search for microlensing the most perspective
objects are gravitational lensed systems as usually, like CLASS
gravitational lens B1600+434, for instance. However, for direct
resolving these images and direct detection of apparent motion of
the knots, a Radioastron sensitivity have to be improved, since an
estimated flux density is too low and to observe the phenomena one
should improve sensitivity in 10 times at 6~cm wavelength, for
instance, otherwise, it is necessary to increase an integration time
(assuming that a radio source has a typical core -- jet structure
and microlensing phenomenon is caused superluminal apparent motion
of knots). In the case of a confirmation (or a disproval) of claims
about microlensing in gravitational lens systems one can speculate
about a microlens contribution into the gravitational lens mass.
Astrometric microlensing due Galactic MACHOs actions is not very
important because of low optical depths and long typical time
scales. Therefore, a launch of space interferometer Radioastron will
give new excellent facilities to investigate microlensing in radio
band, since in this case there is a possibility not only to resolve
microimages but also observe astrometric microlensing.

\end{abstract}


\section{Introduction. Microlensing for distant quasars. }
%\subsection*{\Large Microlensing for distant quasars}

Gravitational microlensing effect was predicted by
\cite{Byalko69,Pacz86} (if sources are stars in Milky Way or Large
Magellanic Cloud discovered by MACHO, EROS and OGLE collaborations
\citep{Macho93,Eros93,Ogle93} discussed in details later in a number
of papers (see, for example,
\cite{Zakharov97,Zakharov98,Zakharov03,Zakharov04b,Kerins01,Griest02,Evans03,Evans03a,Evans04}).
However, microlensing for distant quasars was considered by
\cite{Gott81} (soon after the first gravitational lens discovery by
\cite{Walsh79}) and discovered by \cite{Irwin89} in gravitational
lenses systems since an optical depth for such systems are highest.

 For cosmological locations of gravitational lenses and stellar
masses, typical angles between images are about $\sim 10^{-6}$ sec
\citep{Wamb90,Wamb93,Wamb01}, or more precisely
\begin {eqnarray}
\theta_E =\frac{R_E}{D_S} \approx 2.36 \times 10^{-6}{h_{65}}^{-1/2}
\sqrt{\frac{M}{M_\odot}} \mbox{\rm ~arcsec}, \label{eq0}
\end {eqnarray}
where $R_E$ is the Einstein  -- Chwolson radius, $D_S$ is an angular
diameter distance between a source and an observer,
$h_{65}=\dfrac{H_0}{\mbox{(65 km/(c $\cdot$ Mpc))}}$, $H_0$ is the
Hubble constant.

Theoretical studies of microlensing in gravitational lens systems
started since \cite{Chang79} paper. Unfortunately, till now it is
impossible to resolve microimages, however in this case there is a
chance to observe temporal variations of observed fluxes, or so
called photometric microlensing.

In principle the gravitational lens effect is achromatic, but sizes
and locations for different spectral bands could be different and in
this case we could observe chromatic effect \citep{Wamb_Pacz91}.

\section{Astrometric microlensing}

Astrometric microlensing was discussed in number of papers
\citep{Hog95,Walker95,Miyamoto95,Sazhin96,Sazhin98,Paczynski98,Boden98,Tadros98,Honma01,Honma02,Asada02,Takahashi02,Totani03,Inoue03},
but actually that is signature of well-known light bending in the
gravitational field and at the first time light bending by
gravitational field was discussed by \cite{Newton27}, the first
published derivation of light bending for light was given by
\cite{Soldner04} in the framework of Newtonian theory of
gravitation. In the framework of general relativity light bending
was calculated by \cite{Einstein65} and his prediction was confirmed
in 1919 \citep{Dyson79}. Actually such an astrometric displacement
of distant image due to light bending by gravitational field of
microlenses is called astrometric microlensing and the effect could
be detectable with optical astrometric mission like SIM (Space
Interferometry Mission, see http://sim.jpl.nasa.gov),
   GAIA (Global Astrometric Interferometer for Astrophysics, see http://sci.esa.int/gaia)
and radio projects like VERA (VLBI Exploration
  of Radio Astrometry) and Radioastron.

\subsection{Microlenses in our Galaxy}

Let us remind basic definitions and their relations. We consider a
point size lens. A distance between source and an observer is $
D_s$, a distance between a gravitational lens and observer is $ D_d$
, a distance between a gravitational lens and a source is $ D_{ds}$.
Thus, we obtain gravitational lens equation (Schneider et al. 1992)
\begin{eqnarray}
{\vec{\eta}} ={D_s}{\boldmath{\xi}}/D_d  + {D}_{ds}
\vec{\Theta}(\vec{\xi}),
\end{eqnarray}
where vectors \mbox{\boldmath{$\eta,\xi$}} define coordinates in the
source and lens planes correspondingly, but the angle is determines
by the relation
\begin{eqnarray}
\vec{\Theta}(\vec{\xi})= 4GM \vec{\xi}/{c^2 \xi^2}. \label{eq1_4}
\end{eqnarray}
If the right hand side (\ref{eq1_4}) is equal to zero, we obtain the
conditions when a source, a lens and an observer are located on the
same line (\mbox{\boldmath{$\eta$}=0}). The corresponding length
 $\xi_0= \sqrt{{4GM}{D_dD_{ds}}/({c^2}{D_s})}$
is called Einstein -- Chwolson radius. One could calculate also
Einstein -- Chwolson angle $\theta_0=\xi_0/D_d$.

If we write gravitational lens equation in dimensionless variables,
then we obtain

\begin{eqnarray}
\vec{x} = \vec{\xi}/\xi_0, \quad
 \vec{y} = {D_S} \vec{\eta}/(\xi_0D_d),
\quad
%\nonumber \\
\vec{\alpha} = \vec{\Theta} D_{ds}D_{d}/(D_s\xi_0),
\end{eqnarray}
and the gravitational lens equation has the following form:
\begin{eqnarray}
\vec{y} = \vec{x} - \vec{\alpha}(\vec{x}) \quad \mbox{or} \quad
\vec{y} = \vec{x} - \vec{x}/{x^2}.
\end{eqnarray}
Solving the equation \mbox{\boldmath {$ x$}}, we obtain
\begin{eqnarray}
\vec{x^{\pm}} = \vec{y} [{1}/{2} \pm \sqrt{{1}/{4}+{1}/{y^2}}].
\end{eqnarray}

Then we calculate distance between images:
%\begin{eqnarray}
\begin{gather}
  x^{+} = y\left[\frac{1}{2} + \sqrt{\frac{1}{4}+\frac{1}{y^2}}\right],
\quad
  x^{-} = y\left[-\frac{1}{2} + \sqrt{\frac{1}{4}+\frac{1}{y^2}}\right],
%\nonumber
\notag\\
l=  x^{+}+x^{-} =2 y \sqrt{\frac{1}{4}+\frac{1}{y^2}}.
\label{eq1_11}
\end{gather}
%\end{eqnarray}

\subsection{Typical time scales for astrometric microlensing in our
Galaxy}

Let us consider asymptotic for $x^+$ and $y \rightarrow \infty$,
then $x^+ \rightarrow y+\dfrac{1}{y}$ and angular distance between
real image position  and image position in Einstein -- Chwolson
angles $\Delta = x^+-y \sim \dfrac{1}{y}$ (the angle  describes an
astrometric microlensing).

Let us remind typical scales for lengths, time and angles. Let us
consider the Galactic case if a gravitational lens has stellar mass
$\sim M_{\odot}$  and is located at 10~kpc, then
 \bea
 \xi_0:=
\left[ \left( \frac{ 4GM}{c^2} \right) ~ \left( \frac{ D_d(D_s -
D_d)}{ D_s} \right) \right]
^{1/2} %=
\nonumber\\
= 9.0~\mbox{A.U.} \left( \frac{ M }{ M_{\odot} } \right) ^{1/2}
\left( \frac{ D_d}{10 ~\mbox{kpc}} \right) ^{1/2} \left( 1 -
\frac{D_d}{D_s } \right) ^{1/2} .  \label{mic7} \eea

Thus, we have for Einstein -- Chwolson angle
 \bea \theta_{0} := \left[ \left(
\frac{4GM}{ c^2 } \right) ~
\left( \frac{ D_s - D_d}{D_s D_d } \right) \right] ^{1/2} %=
\nonumber\\
= 0.902 ~ \mbox{mas} ~ \left( \frac{M}{ M_{\odot} } \right) ^{1/2}
\left( \frac{ 10 ~\mbox{kpc}} {D_d } \right) ^{1/2} \left( 1 -
\frac{D_d}{ D_s } \right) ^{1/2} . \label{mic8} \eea

It is known that a distance between images is about $ \sim 2 \xi_0$
for small $y$, thus the angular distance about  (mas). Due to a
proper motion, we have

\bea \dot r = \frac{V}{D_d } = 4.22 ~ \mbox{\rm mas}\cdot ~
\mbox{year}^{-1} ~ \left( \frac{ V }{200~\mbox{km} \cdot
\mbox{c}^{-1} } \right) \left( \frac{10 ~ \mbox{kpc}}{D_d } \right)
, \label{mic9} \eea where $ V $ is a transverse velocity of a lens.
Using last two expressions, one calculates typical time scale for
microlensing, which a time  to cross Einstein radius by a source due
to a proper motion (all distance could be considered at a celestial
sphere):

\bea t_0 := \frac{\theta_0}{ \dot r } = 0.214 ~ \mbox{year} ~ \left(
\frac{ M }{M_{\odot} } \right)^{1/2} \left( \frac{D_d}{10
~ \mbox{kpc}} \right)^{1/2} %\times
\nonumber\\
\times \left( 1 -  \frac{D_d}{D_s } \right)^{1/2} \left( \frac{ 200
~ \mbox{km} \cdot \mbox{c}^{-1}}{V } \right) . \label{mic10} \eea

Let us present rough estimates of an optical depth for astrometric
microlensing using estimates for classic microlensing given by MACHO
and EROS  collaborations $\tau_{\rm halo} \sim 1. \times 10^{-7}$.
Since image displacement for classic microlensing is about
$\theta_{\rm class} \sim 1~mas$, then an optical depth to have
displacement $\theta_{\rm threshold}=10\mu as$ and $\theta_{\rm
threshold}=\mu as$ is given by the expression
\begin {eqnarray}
\tau_{\rm astromet}=\tau_{\rm halo} \left(\frac{\theta_{\rm
class}}{\theta_{\rm threshold}}\right)^2.
 \label{mic16}
\end {eqnarray}
So,   for $\theta_{\rm threshold}=10~\mu as$ an optical depth is
about   $\tau_{\rm astromet} \sim 1. \times 10^{-3}$ and for
$\theta_{\rm threshold}=\mu as$ it is about $\tau_{\rm astromet}
\sim 0.1$, and since according to last estimates $\tau_{\rm halo} =
1.2 \times 10^{-7}$ \citep{Alcock00b,Griest02}. An optical depth for
classical microlensing toward Galactic bulge is about $ \sim 3
\times 10^{-6}$ \citep{Alcock00a}, thus an optical depth for
astrometric microlensing is higher.

We assume that typical time scale for astrometric microlensing is
double time to change an image position displacement from
$\theta_{\rm threshold}$ to maximal displacement $\theta_{\rm max}$.
A typical maximal displacement is $\theta_{\rm
max}=\dfrac{\sqrt{2}}{2}\theta_{\rm threshold}$. Then typical time
scales for astrometric microlensing (one could use other definitions
but difference with the definition could be described by a factor
$\sim 1$)
\begin {eqnarray}
t_{\rm astromet}=t_0 \frac{\theta_{\rm  class}}{\theta_{\rm
threshold}}.
 \label{mic17}
\end {eqnarray}
So,   for $\theta_{\rm threshold}=10~\mu as$ a typical time scale is
about $t_{\rm astromet} \sim 20$~years and for $\theta_{\rm
threshold}=\mu as$ it is about $t_{\rm astromet} \sim 200$~years.

\section{Projected parameters of the space interferometer Radioastron}

According to the schedule of the Russian Space Agency the space
radio telescope Radioastron will be launched in the next few years
(see description of the project  \citep{Kardashev97}).
 This project was initiated by Astro Space Center (ASC)
of the Lebedev Physical Institute of the Russian Academy of Sciences
(RAS) in collaboration with other institutions of RAS and Russian
Space Agency. Scientists from 20 countries develop the scientific
payload for the satellite and will provide a ground base support of
the mission. The project was approved by RAS and Russian Space
Agency. This space based 10-meter radio telescope will be used for
space -- ground VLBI measurements. For observations four wavelength
bands will be used corresponding to $\lambda=1.35$~cm,
$\lambda=6.2$~cm, $\lambda=18$~cm, $\lambda=92$~cm.

It will be not the first attempt to build a telescope with a size
larger than the Earth size. In 1997 Institute of Space and
Technology of Japan launched a HALCA satellite with 8 m radio
telescope and as a result VLBI Space Observatory Programme (VSOP)
was formed \cite{Horiuchi04}. Since the apogee height for
radiotelescope HALCA was 21,200~km, the apogee height for
Radioastron should about 350,000~km (or even $3.5 \times 10^6$~km
see below), and as a result the fringe size for the minimal
wavelength will be smaller than 1-10$\mu as$. The minimal correlated
flux for space-ground VLBI should be about 100 mJy
 for the 1.35~cm wavelength at 8$\sigma$ level \citep{Kardashev97},
therefore source fluxes should be higher than the threshold and
about 24 mJy for the 6~cm wavelength.

An orbit for the  satellite was chosen with high apogee and with
period of satellite rotation around the Earth 9.5 days, which
evolves as a result of weak gravitational perturbations from the
Moon and the Sun. The perigee is in a band  from 10 to 70 thousand
kilometers, the apogee is a band from 310 to 390 thousand
kilometers. The basic orbit parameters will be the following: the
orbital period is p = 9.5 days, the semi-major axis is a = 189 000
km, the eccentricity is e = 0.853, the perigee is H = 29 000 km.

A detailed calculation of the high-apogee evolving orbit can be
done if the exact time of launch is known.

After several years of observations,  it would be possible to move
the spacecraft  to a much higher orbit (with apogee radius about
$3.2\cdot 10^6$~km), by additional spacecraft maneuver using
gravitational force of the Moon. In this case it would be necessary
to use 64-70~m antennas for the spacecraft control, synchronizations
and telemetry.\footnote{http://www.asc.rssi.ru/radioastron/}

 The fringe sizes
(in micro arc seconds) for the apogee of the above-mentioned orbit
and for all Radioastron bands are given in Table \ref{tabl1}.

\begin{table}
\begin{center}
\caption{The fringe sizes (in micro arc seconds) for the standard
and advanced apogees $B_{max}$ (350 000 and 3 200 000~km
correspondingly).}
%\scalebox{1.7}
{
\begin{tabular}{|c|c|c|c|c|}
\hline
$B_{max}({\rm km}) \backslash \lambda ({\rm cm})$ & 92 & 18 & 6.2 & 1.35 \\
\hline \hline
$3.5\times 10^{5}$ & 540 & 106 & 37 & 8 \\
\hline
$3.2 \times 10^{6}$ & 59 & 12 & 4 & 0.9 \\
\hline
\end{tabular}
}
\end{center}
\label{tabl1}
\end{table}

Thus, there are non-negligible chances to observe mirages (shadows)
around the black hole at the Galactic Center and in nearby AGNs in
the radio-band using Radioastron facilities
\citep{ZNDI05,ZNDI05b,Zakharov_Protvino04,ZNDI05d,Zakharov_Texas04,ZDIN05}.

\section{Microimage resolving for distant quasars}
\subsection{Microlens locations}

If microlenses are located in our Galaxy, recent observations by
MACHO, EROS, OGLE collaborations (and their theoretical
interpretations) showed that an optical depth for Galactic microlens
is about $10^{-6}-10^{-7}$. In spite of the fact that for a selected
source a  probability for microlensing is very small and for the
discovery one could monitor about
 $10^6$ background sources (like for microlensing in our Galaxy). That is a hard problem because we have not enough
background point-like distant sources, however an
 angular distance between images is about $10^{-3}$ arcsec,
 therefore there is a possibility to resolve point-like quasar images with VLBI
 technique in radio bands, but unfortunately a sample of bright extragalactic sources is small
to realize the program
  (there is also a chance to resolve the
stellar images in IR band with the modern optical telescopes
\cite{Delpl01,Pacz03}). It was shown that an optical depth for
microlenses located in halo or (and) in quasar bulge is low
\citep{Zakharov04}. We will not study the case because of the
optical depth is low but also angular distance between images is
much shorter than the  Radioastron fringe size.

%%%%%%%%%%%%%%%%%

\subsection{Cosmological distribution for microlenses}

Let us consider cosmologically distributed microlenses since there
is a hypothesis that variability of essential fraction of distant
quasars is caused by microlensing. If it is, one can say that a
probability (an optical depth) is high in radio band also.

To evaluate an optical depth we will assume that a source is located
at a distance with cosmological redshift $z$. Calculations for
different parameters are given by
\cite{Zakharov04,ZPJ_05_IAU225,ZPJ_05_Moriond04}. We will remind of
the results. In calculations we used point-like source approximation
(it means that as a result we obtain an upper limit for an optical
depth).

An optical depth could be evaluated using approximations given by
\cite{Turner84,Fukugita91}
\begin {eqnarray}
\tau^p_L= \frac{3}{2}\frac{\Omega_L}{\lambda(z)} \int_0^z dw
\frac{(1+w)^3[\lambda(z)-\lambda(w)]\lambda(w)}
       {\sqrt{\Omega_0(1+w)^3+\Omega_\Lambda}},
       \label{eq_cosmol2}
\end {eqnarray}
where $\Omega_L$ is compact lens density (in critical density
units), $\Omega_0$ is matter density, $\Omega_\Lambda$ is a
$\Lambda$-term density (or quintessence),
\begin {eqnarray}
\lambda(z)= \int_0^z
\frac{dw}{(1+w)^2\sqrt{\Omega_0(1+w)^3+\Omega_\Lambda}},
\label{eq_cosmol3}
\end {eqnarray}
is an affine parameter (in $cH^{-1}_0$ units).

We use realistic cosmological parameters to evaluate integral
(\ref{eq_cosmol2}). Observations of cosmological SN Ia and CMB
anisotropy give the following parameters $\Omega_\Lambda \approx
0.7, \Omega_0 \approx 0.3$ (or so-called concordance model
parameters). For example, recent observations of the WMAP team gives
for the best fit $\Omega_\Lambda \approx 0.73, \Omega_0 \approx
0.27$ \citep{Bennett03,Spergel03}.

Thus,  $\Omega_0=0.3$ and $\Omega_L = 0.05$ ($\Omega_L = 0.01$)
could be adopted as realistic, if we assume that almost all baryonic
matter form microlenses ($\Omega_L = 0.05$), or 20\% baryonic matter
forms microlenses ($\Omega_L = 0.01$)). However, for $z \sim 2.0$
optical depth could be about  $ \sim 0.01 - 0.1$ \citep{Zakharov04}.
If about 30\% non-baryonic dark matter forms cosmologically
distributed objects with stellar masses (such as neutralino stars
suggested by \cite{Gurevich95,Gurevich96,Gurevich97}, parameter
$\Omega_L = 0.1$ could be adopted as realistic and in this case an
optical depth could be about $ \sim 0.1$. Therefore, if 25\% of
baryonic matter form cosmologically distributed microlenses one
could say that the Hawkins's hypothesis that microlensing cause
variability for essential fraction of all quasars should be ruled
out, but in the case when 30\% of non-baryonic dark matter form
microlenses about 10\% of distant quasars demonstrate these
features.

\subsection{Observed features of microlensing for quasars}

More than 10 years ago
\cite{Hawkins93,Hawkins96,Hawkins02,Hawkins03}
 put
forward the idea that nearly all quasars are being microlensed
(however, based on photometric observations of sample about 25,000
quasars, \cite{Berk04} claimed that microlensing model for an
explanation of variabilities is unlikely).

As previous estimates show us that in the case if Hawkins hypothesis
is correct, $\Omega_L$ should be about 1 and that is a contradiction
for data of observational cosmology, but hys hypothesis could be
correct in part and if observations would dictate that $\Omega_L$ is
larger than 0.05 we could conclude that non-baryonic matter form
microlenses (they could be neutralino clouds or primordial black
holes). If the Hawkins hypothesis is correct in part at least, in
this case also an essential fraction of distant point like sources
should demonstrate features of microlensing since the optical depth
could be evaluated by Eq. (\ref{eq_cosmol2}) as well. No doubt that
except microlensing there are other causes of variabilities, however
one could use different techniques to separate  different types of
variabilities (see, \cite{Koopmans2000,Koopmans2000b}, for example),
since there is different dependence of modulation indices as a
function of frequency for oscillations (scintillations) and for
microlensing. However, resolving the microimages and measuring the
centroid displacements for bright point-like sources in radio band
will be a critical test to prove (or rule out) the Hawkins
hypothesis about microlensing for point like sources at cosmological
distances.

To prove the microlensing hypothesis for a distant quasar, the
source have to have the following properties from  a list of perspective targets
of VSOP or Radioastron missions (or from its extended version): \\
a) A source should demonstrate signatures of microlensing which are
different from typical features for scintillations at time scales
$<$ 3--5 years
(that is an estimated time of Radioastron mission);\\
b) A compact core for the source should have size $\lesssim 40 \mu
as$ and flux density should be higher than Radioastron thresholds
$\gtrsim 20$ mJy at 6~cm wavelength and  $\gtrsim 100$ mJy at at
1.35~cm wavelength.

In the case, if the Hawkins hypothesis is correct an essential
fraction of all point like sources at cosmological distances should
demonstrate signatures of photometric and therefore astrometric
microlensing.

In the case, if the Hawkins hypothesis is incorrect and
cosmologically distributed microlenses give a small contribution
into critical density $\Omega_{\rm tot}$, but even for this case one
could evaluate  $\Omega_L$ from an observed rate  of microlensed
sources satisfying condition b), since the observed rate gives an
estimate for the optical depth.

According to \cite{Horiuchi04} results about $14\% \pm 6\%$ of
sources (from 344 ones) have core size $\lesssim 40 \mu as$ and the
angle corresponds to the fringe size at the 6~cm wavelength. This
part of sources could be used for photometric monitoring and for a
further analysis of a preferable explanation of variability. If the
analysis would indicate that microlensing is a preferable cause of
variability the candidate could be selected as the first order one.
But even in the case, if a source would demonstrate variability that
could be explained by another cause (but not by microlensing), the
source should be checked to search for image splitting or (and)
astrometric image displacement since models for alternative
explanation of variabilities could be not quit correct.

From theoretical point of view there is a possibility to detect
microlensing for both core and bright knots. In this case the two
situations will be characterized by different time scales.

%Observations with space interferometer Radioastron appearance or
%disappearance of micro images could confirm or rule out Hawkins
%hypothesis that variability of a large part of quasars is caused
%by microlensing, since if Radioastron will have a high orbit than
%it will have an angular resolution about  $10^{-6}$ arcsec.

\section{Microlensing for gravitational lensed systems}

Few years ago, \cite{Koopmans2000,Koopmans2000a} claimed that the
most realistic explanation of short-term variability of a
gravitational lens CLASS B1600+434 at 5~GHz and 8.5~GHz
(variabilities and possible explanations of the phenomena were
discussed by \cite{Koopmans03,Winn04}). The authors considered
different cases of variability such as scintillation due to
scattering and microlensing. As a result they concluded, that
microlensing phenomenon in radio band gives the natural fit for
observational data. One could remind flux densities changed from
58(48) mJy in March 1994 to 29 (24) mJy in August 1995 for image
A(B) \citep{Koopmans98}. Another decrease was found from 27(24) to
23(19) mJy and it was from February to October 1998
\cite{Koopmans2000b,Koopmans2000}. Strong variability was detected
at 5~GHz, where flux density was about 34--37 mJy in 1987
\cite{Becker91,Koopmans2000}, but it was about 45(37) mJy for image
A(B) \citep{Koopmans98} and only 23 (18) mJy in June 1999
\citep{Koopmans2000}.
 Based on analysis of variabilities
\cite{Koopmans2000} concluded that the variability is caused by
superluminal motion of compact knots in jet (VLBA and 100-m
Effelsberg telescope observations also found evidences for jet
components in the CLASS gravitational lens B0128+437
\citep{Biggs04}, but unfortunately their flux densities are too low
to observe then with the Radioastron interferometer).

Let us remind that a typical threshold for Radioastron
interferometer sensitivity at 5~GHz is about 23 mJy with  an
integration time 300~s \citep{Kardashev97}, therefore in principle,
such density fluxes could be detected by Radioastron interferometer.

\cite{Treyer03} concluded that for photometric fluctuations $\sim
0.5$~mag typical astrometric displacement should be about 20 --
40~$\mu as$ (to evaluate photometric and astrometric microlensing
one could use numerical approaches and analytical asymptotical
expansions near fold \citep{Schneider92a} and cusp singularities
\citep{Zakharov95,Zakharov97,Petters01,Yonehara01}. In principle
such a displacement could be observed with Radioastron space
interferometer at 6~cm and 1.35~cm wavelengths if flux densities for
the object is high enough. For example, in the B1600+434 case the
density flux is suitable for the core (at least, at 6~cm
wavelength), but if the superluminal motion of knots is responsible
for microlensing (as \cite{Koopmans2000} claimed) the sensitivity of
Radioastron should be improved in 10 times at 6~cm wavelength to
observe such a displacement of knots. At the 1.35~cm wavelength the
Radioastron flux density threshold is probably too high to detect
the displacement.

\subsection{Typical time scales for microlensing}

Let us remind that according to the standard  model typical time
scales for radio microlensing could be much smaller than typical
time scale in optical band due to effects of special relativity and
different geometry and locations of radiating regions  in these
bands, for example typical time scales in optical band are
determined by a transverse velocity ($v_{\rm trans}$), but in
radioband time scales could be in $\beta_{\rm trans}/v_{\rm trans}$
times smaller \citep{Koopmans2000} (all velocity are expressed in
$c$ units).

Typical time scales is determined by a ratio typical sizes between
caustics and an apparent velocity of the jet-component in the source
plane \citep{Blandford77,Blandford79,Koopmans2000}. If jet-component
moves with a relativistic bulk velocity ${\bf{\beta}}_{\rm bulk}$,
then an apparent velocity ${\bf{\beta}}_{\rm app}$ is
\begin {eqnarray}
{\bf{\beta}}_{\rm app}=\frac{\bf{n} \times ({\bf{\beta}}_{\rm bulk}
\times \bf{n})}{1- {\bf{\beta}}_{\rm bulk} \cdot \bf{n}} =
\frac{{\bf{\beta}}_{\rm bulk} \sin (\psi)}{1- |{\bf{\beta}}_{\rm
bulk}| \cos(\psi)} , \label{eq_time_new_1}
\end {eqnarray}
where $\psi$ is the angle between the jet and a line of sight
\citep{Blandford77,Blandford79,Koopmans2000}.

The apparent angular velocity of the jet component is
\citep{Koopmans2000}
\begin {eqnarray}
\frac{d{\bf{\theta}}_{\rm s}}{dt}= \frac{{\bf{\beta}}_{\rm app}
c}{(1+z_{\rm s})D_{\rm s}} = \frac{1.2 \cdot {\bf{\beta}}_{\rm
app}}{(1+z_{\rm s})}\frac{\rm Gpc}{D_{\rm s}} \frac{\rm \mu as}{\rm
week},  \label{eq_time_new_2}
\end {eqnarray}
where $z_{\rm s}$ and $D_{\rm s}$ are the source redshift and the
angular diameter distance to the stationary core, respectively.
Using the estimate for observed source redshift $z_{\rm s}$
\citep{Fassnacht98}, \cite{Koopmans2000} concluded that angular
velocity of B1600+434 should be
\begin {eqnarray}
\frac{d{\bf{\theta}}_{\rm s}}{dt}= 0.34 \cdot {\bf{\beta}}_{\rm app}
\frac{\rm \mu as}{\rm week}, \label{eq_time_new_3}
\end {eqnarray}
for a flat Friedmann universe with $\Omega_{\rm m}=1$ and
$H_0=65~{\rm km}\cdot {\rm s}^{-1}{\rm Mpc}^{-1}$. Based on
observational data and simulations \cite{Koopmans2000} evaluated
also a typical size of knots in jet in the source plane $2< \Delta
\theta_{\rm knot} <5~{\rm \mu as}$ and an apparent velocity band $9<
\Delta \theta_{\rm app} <26$.\footnote{However, probably explanation
of variabilities by superluminal motions in jet and microlensing
should be verified, because \cite{Patnaik01} discussed observations
which are in contradiction with the model \cite{Koopmans2000}.}
Therefore, apparent displacements for B1600+434 should be about
about dozens $\mu as$ and the displacement could be measured with
the Radioastron interferometer at 6~cm wavelength.

One could also evaluate linear sizes of knots through their angular
diameter distances
\begin {eqnarray}
\Delta ~l = \frac{c}{H_0}\frac{\Delta \theta_{\rm knot}
\left[z_s-(1+q_0)z_s^2/2\right]}{1+z_s}, \label{eq_time_new_3a}
\end {eqnarray}
where $q_0=1.3\cdot\Omega_m -1 = -0.55$ (for a flat universe and
$\Omega_m=0.3$), therefore typical linear sizes of the knots should
be $\Delta l \in (5, 14)10^{16}$~cm.

Typical scales for microlensing are discussed not only in books on
gravitational lensing (Schneider et al. 1992, Petters et al. 2001),
but in recent papers also (see, for example, \cite{Treyer03}).
Usually people discuss  locations of microlenses in gravitational
macrolenses because of an optical depth for microlensing is the
highest in comparison with other possible locations of gravitational
microlenses,  but it is clear that the fact it was known quit well
in advance. However, cases for microlens locations were considered,
for example galactic clusters or extragalactic dark halos could have
microlenses.

So, for example following to a recent paper by \cite{Treyer03}, we
remind that typical length scale for microlensing and assuming
concordance cosmological model parameters ($\Omega_{\rm tot}=1,
\Omega_{\rm matter}=0.3, \Omega_{\Lambda}=0.7$)
\begin {eqnarray}
 R_E = \sqrt{2 r_g \cdot  \frac{D_s D_{ls}}{D_{l}}}
 \approx 3.4 \cdot 10^{16} \sqrt{\frac{M}{M_\odot}} h_{65}^{-0.5}{\rm cm},
 \label{eq_suppl1}
\end {eqnarray}
where "typical" microlens and sources redshifts are assumed to be
$z_l=0.5, z_s=2$ (similar to \cite{Treyer03}),
$r_g=\dfrac{2GM}{c^2}$ is the Schwarzschild radius corresponding to
microlens mass $M$, $h_{65}=H_0/((65~{\rm~km/sec})/{\rm Mpc})$ is
the dimensionless Hubble constant.

The corresponding angular scale is \citep{Treyer03}
\begin {eqnarray}
\theta_0=\frac{R_E}{D_s}
 \approx 2.36 \cdot 10^{-6} \sqrt{\frac{M}{M_\odot}} h_{65}^{-0.5}{\rm ~arcsec},
 \label{eq_suppl2}
\end {eqnarray}

Using the length scale (\ref{eq_suppl1}) and a velocity scale (say
an apparent velocity $\beta_{\rm app}$), one could calculate the
standard time scale corresponding to the scale to cross the Einstein
radius
\begin {eqnarray}
t_E=(1+z_l)\frac{R_E}{v_\bot}=%\\
\begin{cases}
 \approx 2 \sqrt{\dfrac{M}{M_\odot}}{\beta_{\rm app}}^{-1} h_{65}^{-0.5}{\rm
 ~weeks}, & \text{if $v_\bot=c\beta_{\rm app},$}\\
\approx 27 \sqrt{\dfrac{M}{M_\odot}}{v_{600}}^{-1} h_{65}^{-0.5}{\rm
 ~years}, & \text{if $v_\bot \sim 600~{\rm km/c},$}\\
\end{cases}
 \label{eq_suppl3}
\end {eqnarray}
here we assume time scales are determined by an apparent velocity or
a typical transverse velocity ($v_{600}=v_\bot /(600~{\rm km/c})$),
respectively.

The time scale $t_E$ corresponding to the approximation of a point
mass lens and small size of source in comparison with Einstein --
Chwolson radius and probably the approximation and the time scale
could be used if microlenses are distributed freely at cosmological
distances and actually one Einstein -- Chwolson angle is located far
enough from another one.

If we use the simple caustic microlens model (like the straight
fold caustic model), there are two time scales, namely it depends
on sizes of "caustic size" and  source radius $R$. If the source
radius is larger or about "caustic size" $r_{\rm caustic}$ (if we
use the following approximation for the magnification near the
caustic $\mu=\sqrt{\dfrac{r_{\rm caustic}}{y-y_c}}$ ($y>y_c$ and
$y$ is the perpendicular direction to the fold caustic)), thus $R
\gtrsim r_{\rm caustic}$, then the relevant time scale is the
"crossing caustic time"  \citep{Treyer03}
\begin {eqnarray}
t_{\rm cross}=(1+z_l)\frac{R_{\rm source}}{v_\bot (D_s/D_l)} \nonumber\\
 \approx 0.62\ R_{15} v_{600}^{-1} h_{65}^{-0.5}{\rm ~years}\nonumber\\
\approx 226\ R_{15} v_{600}^{-1} h_{65}^{-0.5}{\rm ~days},
 \label{eq_suppl4}
\end {eqnarray}
(in the right hand side $D_l$ and $D_s$ correspond to $z_l=0.5$ and
$z_s=2$ respectively and $R_{15}=R_{\rm source}/10^{15}$~cm).

However, if the source radius $R_{\rm source}$ is much smaller
than the "caustic size" $r_{\rm caustic}$  $R_{\rm source} \ll
r_{\rm caustic}$, one could used the "caustic time", namely the
time when the source is located in the area near the caustic and
the time scale corresponds to
\begin {eqnarray}
t_{\rm caustic}=(1+z_l)\frac{r_{\rm caustic}}{v_\bot (D_s/D_l)} \nonumber\\
 \approx 0.62\ r_{15} v_{600}^{-1} h_{65}^{-0.5}{\rm ~years}\nonumber\\
\approx 226\ r_{15} v_{600}^{-1} h_{65}^{-0.5}{\rm ~days},
 \label{eq_suppl5}
\end {eqnarray}
where $r_{15}=r_{\rm caustic}/10^{15}$~cm.

These time scales $t_{\rm cross}$ and $t_{\rm caustic}$ could be
about days (or even hours) if $v_\bot$ is determined by an apparent
motion of superluminal motion in jet.

Thus $t_{\rm cross}$ could be used as a lower limit for typical
time scales for the simple caustic microlens model, but since
there are two length parameters in the problem and in general we
do not know their values, we could not evaluate $R_{\rm source}$
only from the time scales of microlensing because time scales
could correspond to two different length scales. However, if we
take into account variation amplitudes of luminosity, one could
say that in general $t_{\rm cross}$ corresponds to to smaller
variation amplitudes than $t_{\rm caustic}$, because if the source
square is large there is a "smoothness" effect since only small
fraction of source square is located in the high amplification
region near the caustic.

\section{Conclusions}

First, one could point out that gravitational lensed systems are the
most perspective objects to search for microlensing. Astrometric
microlensing could be detected in the gravitational lens system such
as B1600+434 in the case if a proper motion of source, lens and an
observer are generated mostly by a superluminal motion of knots in
jet (superluminal motion in jet was found with HALCA in the quasar
PKS~1622-297 \citep{Wajima05}). But in this case, based on density
flux estimates done by \cite{Koopmans2000}, one could say that a
required
 sensitivity of the Radioastron interferometer should be improved in
 10 times.

In the case if there is microlensing of core in the B1600+434 system
for example, then astrometric microlensing in the system could be
about should be about 20 -- 40~$\mu as$  \citep{Treyer03} and the
Radioastron interferometer will have enough sensitivity to detect
such an astrometric displacement.

Second, in principle microlensing for distant sources could be the
only tool to evaluate $\Omega_L$ from microlensing event rate.  To
solve this problem with the Radioastron interferometer one should
analyze variabilities of compact sources with a core size $\lesssim
40 \mu as$ and with high enough flux densities about $\gtrsim 20$
mJy at 6~cm wavelength and  about $\gtrsim 100$ mJy at at 1.35~cm
wavelength To fit the most reliable model for variabilities  of the
sources such as scintillations, microlensing etc. A fraction of the
sources in the list of extragalactic targets for VSOP and
Radioastron about 13\% -- 14 \%
\citep{Moellenbrock96,Hirabayashi00,Scott04,Kovalev05}. In the case,
if the analysis would indicate that other explanations (such as
scintillations) are preferable and future observations with
Radioastron interferometer would show that the are no features for
astrometric microlensing, one could conclude that Hawkins hypothesis
should be ruled out. But if an essential fraction of variability
could be fitted by microlensing, the sources could be as the first
order candidate to search for astrometric microlensing.

Therefore, one could say that astrometric microlensing (or direct
image resolution with Radioastron interferometer) is the crucial
test to confirm (or  rule out) microlens hypothesis for
gravitational lensed systems and for point like distant objects.

Astrometric microlensing due to MACHO action in our Galaxy is not
very important for observations with the space interferometer
Radioastron, since first, probabilities are not high; second,
typical time scales are longer than estimated life time of the
Radioastron space mission.

Therefore, just after the Radioastron launch it will be the first
chance to detect microlensing by a direct way. So, the main goal of
the paper to attract an attention to such a challenging possibility
because, preflight time is very short now and perspective targets
should be analyzed carefully by observational and theoretical ways
in advance. A number of point like bright sources at cosmological
distances and gravitational lensed systems with point like
components demonstrating microlens signatures is not very high and
the sources should be analyzed by the careful way to search for
candidates where microlens model is preferable in comparison with
alternative explanations of variabilities.

\section*{Acknowledgements}

The author thanks N.S.~Kardashev for fruitful discussions and an
anonymous referee for critical remarks.
 The author is also grateful to the National Natural Science
Foundation of China (NNSFC)  (Grant \# 10233050) and National Key
Basic Research Foundation (Grant \# TG 2000078404) for a partial
financial support of the work.

%\newpage

\end{document}